\documentclass[aps,pre,amsmath,amssymb,floatfix,twocolumn,showpacs]{revtex4}

\usepackage{graphicx}
\usepackage{dcolumn}

\newcommand{\e}{\mathrm{e}}
\renewcommand{\d}{\mathrm{d}}

\newcommand{\f}{_\mathrm{f}}
\newcommand{\crit}{^\mathrm{max}}
\renewcommand{\P}{L}
\newcommand{\Pb}{P_\mathrm{b}}
\newcommand{\Pnc}{P_\mathrm{nc}}
\newcommand{\Nf}{N_\mathrm{f}}
\newcommand{\Prob}{P}
\newcommand{\Probd}{p}
\newcommand{\pbimodal}{{p_0}}

\begin{document}

\title{Stochastic Load-Redistribution Model for Cascading Failure Propagation}

\author{J\"org Lehmann}
\affiliation{ABB Switzerland Ltd., Corporate Research,
Segelhofstrasse 1K, 
CH-5405~Baden-D\"attwil, Switzerland}
\author{Jakob Bernasconi}
\affiliation{ABB Switzerland Ltd., Corporate Research,
Segelhofstrasse 1K, 
CH-5405~Baden-D\"attwil, Switzerland}

\date{\today}

\begin{abstract}
  A new class of probabilistic models for cascading failure
  propagation in interconnected systems is proposed. The models are
  able to represent important physical characteristics of realistic
  load-redistribution mechanisms, e.g., that the load increments after
  a failure depend on the load of the failing element and that they
  may be distributed non-uniformly among the remaining elements. In
  the limit of large system sizes, the models are solved analytically
  in terms of generalized branching processes, and the failure
  propagation properties of a prototype example are analyzed in
  detail.
\end{abstract}


\maketitle

\section{Introduction}

The increasing complexity of todays infrastructure networks, e.g.,
electrical power grids, road systems, or communication networks, makes
them very sensitive to local
failures~\cite{Motter2002,Watts2002,Dobson2007,Simonsen2008,Buldyrev2009}.
When an element in such a network fails, its ``load'' (e.g., power,
traffic, or information flow) is redistributed to the other elements of
the network. Some of the increased loads may then exceed the capacity
of their respective element, leading to further failures and eventually
to a cascading breakdown of the entire network. Cascading failure
propagation is not only observed in physical infrastructure networks,
but also in social and economic systems~\cite{Motter2002,Watts2002} or
in the fracture of heterogeneous materials~\cite{Alava2006,Pradhan2008}.

As a breakdown of critical infrastructure networks can have serious
economic consequences, it is crucial to gain a deeper understanding of
the mechanisms that lead to such cascading failures.  This problem
has, in particular, attracted the interest of the statistical physics
community, and various models have been developed to study the
vulnerability of complex networks with respect to cascading failure
propagation~\cite{Motter2002,Watts2002,Dobson2007,Bakke2006,Simonsen2008}.
A description of the load-redistribution on different levels of detail
has been considered, e.g., more physical approaches based on resistor
networks~\cite{Bakke2006} or complex-network models focusing on purely
topological measures like the betweenness
centrality~\cite{Motter2002,Albert2004,Huang2008}. The dynamics of
most of these models, however, can only be analyzed via large-scale
numerical simulations. In order to obtain an analytically solvable
model, Dobson et al.~\cite{Dobson2007, Dobson2005} consider the
simplifying assumption that the load increments after a failure are
the same for all remaining elements and independent of the failing
load.  Similarly, fiber bundle models for the problem of fracture
propagation~\cite{Daniels1945,Alava2006,Pradhan2008,Hidalgo2002} can only be solved
analytically if the load of the failing fiber is equally redistributed
to all remaining fibers.

In this paper, we introduce and analyze a new class of probabilistic
models for cascading failure propagation that can represent, in a
stochastic sense, important characteristics of realistic
load-redistribution mechanisms: The load redistribution after a
failure is no longer assumed to be uniform and the induced load
increments may depend on the load of the failing element.  With such
models, we can thus expect to obtain a better understanding of the
breakdown processes in real networks.  We show that in the limit of
large system sizes, our models can be solved analytically by using a
Markov approximation and the theory of generalized branching
processes~\cite{Harris1963}. We then apply our general approach to an
illustrative prototype system that roughly imitates failure
propagation in a power transmission network and analyze its vulnerability with respect
to cascading breakdown.

\section{Cascading-failure model}

We consider a system consisting of
$N$~elements, each with a random load $\P\ge0$. The loads
are assumed to be independent of each other and identically
distributed.  Furthermore, every element possesses a random critical
load $\P\crit$ above which it will fail. Whereas we assume that the
critical loads of the various elements are independent of each other,
we allow for possible correlations between the initial and
critical loads of a particular element. Specifically, we require that
initially none of the elements is overloaded, i.e.,  the probability $\Prob(\P > \P\crit)$ vanishes.

We now consider a situation where, due to some external influence, one
of the elements, say with load~$\P\f$, fails. Our central model
assumption is that this load is redistributed to the remaining
elements according to the \textit{stochastic load-redistribution} rule
\begin{equation}
  \label{eq:redist_1}
  \P \rightarrow \P' = \P +  \P\f \, \Delta\,.
\end{equation}
Here, $\P$ ($\P'$) is the load of one of the remaining elements before
(after) failure of the element with load~$\P\f$, and the
load-redistribution factor~$\Delta$ is a random number drawn
independently from the same distribution for each of the
remaining elements. In other words: The load increments are
\textit{proportional} to the failed load~$\P\f$, but with random
proportionality factors~$\Delta$.

The form of rule~\eqref{eq:redist_1} is based on the observation that
in many systems, the load-redistribution factors primarily depend on
structural properties, such as interactions between the various
elements, and not on the load of the failing element.  In a more
``microscopic'' approach, the failure dynamics of such systems would
be described by a model of the form~\eqref{eq:redist_1}, but with the
factor~$\Delta$ being determined by the specific interactions of the
failing element with the one affected by the failure.  Corresponding
examples range from the power-flow redistribution after a line failure
in power grids~\cite{Baldick2003} to the distance-dependent stress
redistribution in fiber bundles~\cite{Hidalgo2002}. The main features
of a load redistribution of the form~\eqref{eq:redist_1} can already
be understood by considering the extreme cases of a uniform, global
load redistribution and a purely local one.  In the former case, each
element is affected in the same way and thus $\Delta=1/(N-1)$. The
latter situation is described by $\Delta=1/Z$ for the $Z$ nearest
neighbors of the failing element and zero otherwise. The stochastic
load redistribution rule~\eqref{eq:redist_1} models the microscopic
$\Delta$-dependence in terms of a noisy dynamics that neglects any
spatial correlations. While its specific form thus depends on the
system at hand---we will consider an example in
Sect.~\ref{sec:example} below---we expect two properties to be
generally fulfilled: (i) On average, the failed load will be
\textit{redistributed} to the remaining $N-1$ elements. This implies
that the mean $\langle \Delta \rangle$ behaves as $1/N$ for large
$N$. (ii) The $\Delta$-distribution typically will be bounded. For
instance, if---in the worst case---one single element has to take over
the load of the failing element, one has $|\Delta| \le 1$.

So far, we have only discussed the load redistribution after an
initial failure.  Obviously, it can happen that the post-failure loads
of a number~$\Nf^{(1)}\ge1$ of elements are above their respective
critical loads. In such a situation, a failure cascade develops.  For
its description, we assume that the overloaded elements fail
\textit{simultaneously} and that each failing load is redistributed to
the remaining elements according to
rule~\eqref{eq:redist_1}~\footnote{Note that, in general, the
  distribution of the load-redistributions factors~$\Delta$ will
  change as the number of intact elements decreases. How to take into
  account this finite-size effect depends on the system considered. In
  the model~\eqref{eq:bimodal} below, we keep $\Delta_0$ fixed and
  adjust $\pbimodal=1/[(N_\mathrm{nf}-1)\Delta_0]$, where $N_\mathrm{nf}$ is
  the number of intact elements before the new failure.}. If this
redistribution results in further overloading, the cascade continues
to a new cascade stage. This process continues until the system either
reaches a stable state, i.e., the remaining elements operate within
their bounds, or all~$N$ elements have failed and the system has
broken down completely.

Denoting the number of failures at each cascade stage~$s=1,2,\dots$ by
$\Nf^{(s)}$ and counting the initial failure as $N_\mathrm{f}^{(0)}=1$,
the total number of failed elements $\Nf=\sum_{s=0} \Nf^{(s)}$
provides a measure for the damage to the system.  The distribution of
this random variable characterizes the system stability. Coarsely, two
regimes can be distinguished: (i) The probability of large
$N_\mathrm{f}$ decays quickly, i.e., at least exponentially, and thus
system-wide cascades with $\Nf\lesssim N$ constitute very rare events;
(ii) System-wide failures occur with finite probability even for
$N\to\infty$. At the separation between these two regimes, the system exhibits a
``critical'' behavior~\cite{Dobson2007}, where large-scale events are still suppressed
but their probability only decays according to a power law:
$\Prob(\Nf) \propto \Nf^{-\gamma}$ for $N\to\infty$.

To determine the stability of a given system with respect to cascading
failures, the detailed form of the probability
distribution~$\Prob(\Nf)$ is not required and will not be evaluated in
the present paper. Instead, it suffices to have an indicator for the
two regimes just outlined. An obvious choice is the probability for a
system-wide breakdown: $\Pb = \Prob(N_\mathrm{f} = N)$.  Another
quantity of interest is the probability that an initial failure does
not induce any further failures, in other words, the probability that
no cascade develops at all: $\Pnc = \Prob(N_\mathrm{f} = 1)$

\section{Generalized-branching-process approximation}

In the limit of large
systems, $N\to\infty$, when finite-size effects do not play a role, an
approximate description of the cascade dynamics can be obtained by
making two observations: First, during a failure cascade, the
distribution of the not yet failed loads can be approximated by their
initial distribution. Thus, a Markovian description in terms of the
loads which fail at every cascade step becomes possible. The
corresponding states form a point process on the non-negative real
axis~\cite{van_Kampen2007}. Second, as the number of remaining
elements always stays infinitely large, the number of induced failures
can be described by a Poisson distribution. This yields an
approximation of our model in terms of a generalized branching
process~\cite{Harris1963}, which is fully defined by its
characteristic functional
\begin{equation}
  \begin{split}
  G[u;\P\f] = \exp\bigg\{\mu\f(\P\f)\bigg[ \int\! & \d\P\f' \, \, \Probd(\P\f'|\P\f'> \P\crit;\P\f) \\
  & \times\e^{-u(\P\f')} - 1\bigg]\bigg\}\,,
  \end{split}
  \label{eq:chfunc}
\end{equation}
where $u$ denotes an arbitrary non-negative test function on the interval $[0,\infty)$ and
$\Probd(\P\f'|\P\f'> \P\crit;\P\f)$ is the conditional
probability density that a failure induced by a failing load $\P\f$ occurs
with a load~$\P\f'$. Given the joint distribution of initial and
critical loads, as well as the distribution of the load-redistribution factors,
this quantity can be readily calculated.  The mean number
of induced failures is given by $\mu\f(\P\f) = (N-1)\, \Prob(\P\f'>
\P\crit|\P\f)$. Note that in order for a meaningful limit $N\to\infty$
to exist this implies that the conditional failure probability of a single element
$\Prob(\P\f'> \P\crit|\P\f)$ has to be
of order~$O(1/N)$ (cf.\ the discussion above on the mean of
the load-redistribution factors). 

For the calculation of the breakdown and no-cascade probabilities, we condition these quantities on the load~$\P\f$ of
the failing element. From the Poissonian distribution of the failures
induced directly by this initial failure, one
then obtains the conditional no-cascade probability
$\Pnc(\P\f) = \exp[-\mu\f(\P\f)]$. The conditional breakdown
probability~$\Pb(\P\f)$ can be obtained as solution of the integral
equation~\cite{Harris1963}
\begin{multline}
  \label{eq:pb}
  1-\Pb(\P\f) = \exp\bigg\{-\mu\f(\P\f)\int\!  \d\P\f' \, \, \Probd(\P\f'|\P\f'> \P\crit;\P\f)
  \\\times \Pb(\P\f')\bigg\}\,.
\end{multline}
This relation can be interpreted in the sense that the probability that
no complete breakdown develops after a failure with load ~$\P\f$ equals the probability
that---in the limit $N\to\infty$---none of the induced failures with
load $\P\f'$ leads to a breakdown. Starting from an initial guess
for~$\Pb(\P\f)$, Eq.~\eqref{eq:pb} can be efficiently solved by means
of an iterative procedure~\cite{Harris1963}. This either yields the vanishing solution
$\Pb(\P\f)\equiv 0$ if the system is immune against cascading failures or the
unique nontrivial solution with finite breakdown probability. Note that
the range of possible $\P\f$ in Eq.~\eqref{eq:pb} might be larger than
that of the initial loads~$\P$ (see the example in Sect.~\ref{sec:example} below). We finally remark that
an integral relation similar to Eq.~\eqref{eq:pb} can be derived for
the generating function of the total number of failures~$N\f(\P\f)$, where $\P\f$ is the initially failing load.

\section{Example: Simple bimodal load redistribution}

\label{sec:example}

As a simple, yet prototypical example, we now consider
a \textit{bimodal} load redistribution: 
\begin{equation}
  \Delta = 
  \begin{cases}
    \Delta_0 & \text{with probability $\pbimodal$} \\
    0 & \text{with probability $1-\pbimodal$}\,.
  \end{cases}
  \label{eq:bimodal}
\end{equation}
Thus, a failure affects a specific other element with probability~$\pbimodal$,
in which case this element receives a portion~$\Delta_0$ of the failed load.  In
line with the above arguments, we require $\langle \Delta \rangle = \pbimodal
\, \Delta_0= 1/(N-1)$ and consequently $1/(N-1)\le\Delta_0\le 1$. 
On average, the failing load is redistributed to
$(N-1) \pbimodal = 1/\Delta_0$ other elements. In this sense, the model allows one to
study the transition between the above-mentioned two extreme cases of
a global load redistribution for $\Delta_0 = 1/(N-1)$ and a load transfer to a
single other element, typically the nearest neighbor, for
$\Delta_0=1$.

For the initial loads~$\P$, we consider a uniform distribution, which
can be scaled without loss of generality to the interval $[0, 1]$.
Motivated by applications to infra\/structure networks with
cost-limited capacity, e.g., power transmission networks, we assume
that the maximal load of each element is higher than its initial load
by a constant tolerance~$\alpha\ge0$~\cite{Motter2002}:
\begin{equation}
  \label{eq:Lcrit}
  \P\crit = (1 + \alpha)\, \P\,.
\end{equation}
In the limit $N\to\infty$, the model is thus fully characterized by
the two parameters $\Delta_0$ and $\alpha$ and in the following, we shall study
the stability of the system as a function of these parameters.

As shown above, the no-cascade probability~$\Pnc$ follows directly
from the mean number of failures induced by a failure with load
$\P\f$. From $\Prob(\P\f'>\P\crit| \P\f) = \pbimodal \,\Prob( \P < \P\f \,
\Delta_0/\alpha )$, where $\P$ is the initial load of an arbitrary
element, we obtain $\mu\f(\P\f) = \min(1/\Delta_0, \P\f/\alpha)$.  The
integral equation~\eqref{eq:pb} for the conditonal breakdown probability~$\Pb(\P\f)$ assumes the form
\begin{multline}
  \label{eq:pb_ex}
  1-\Pb(\P\f) = \exp\bigg\{-\frac{1}{\Delta_0}\!\!\!\!\int\limits_{\P\f \Delta_0}^{\P\f \Delta_0 + \min(1, \P\f\Delta_0/\alpha)} \!\!\!\!\!\!\!\!\!\!\!\!\!\!\!\! \!\!\!\!\!\d\P\f'\,\,
  \Pb(\P\f')\bigg\}\,.
\end{multline}
It has to be solved on the interval $[0,\P_\mathrm{f,max}]$ with
$\P_\mathrm{f,max}=1/(1-\Delta_0)$ for $\alpha<\Delta_0/(1-\Delta_0)$
and $\P_\mathrm{f,max}=1$ otherwise. 

If we assume that the initially failing element is chosen at random
with equal probability, we obtain the total no-cascade and breakdown
probabilities, $\Pnc$ and $\Pb$, respectively, by integrating the
corresponding conditioned probabilities over the range $[0,1]$ of
possible initially failing loads. Whereas for the breakdown probability,
the integral has to be performed numerically from the iterative
solution of Eq.~\eqref{eq:pb_ex}, the no-cascade probability can be
obtained explicitly:
\begin{equation}
  \label{eq:Pnc}
 \Pnc = 
\begin{cases}\alpha +
(1-\alpha-\alpha/\Delta_0)\, \mathrm{e}^{-1/\Delta_0} &\!\!\text{for $\alpha<\Delta_0$}\\
\alpha-\alpha\, \mathrm{e}^{-1/\alpha} &\!\!\text{for $\alpha\ge\Delta_0$}. 
\end{cases}
\end{equation}

\begin{figure}[ht]
  \centering
  \includegraphics[width=0.98\linewidth]{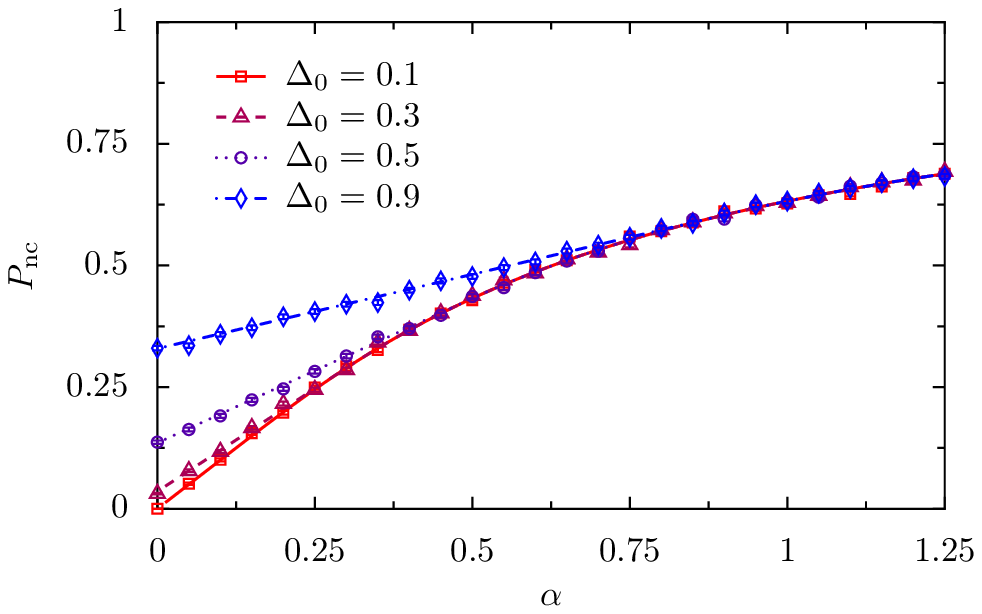}
  \includegraphics[width=0.98\linewidth]{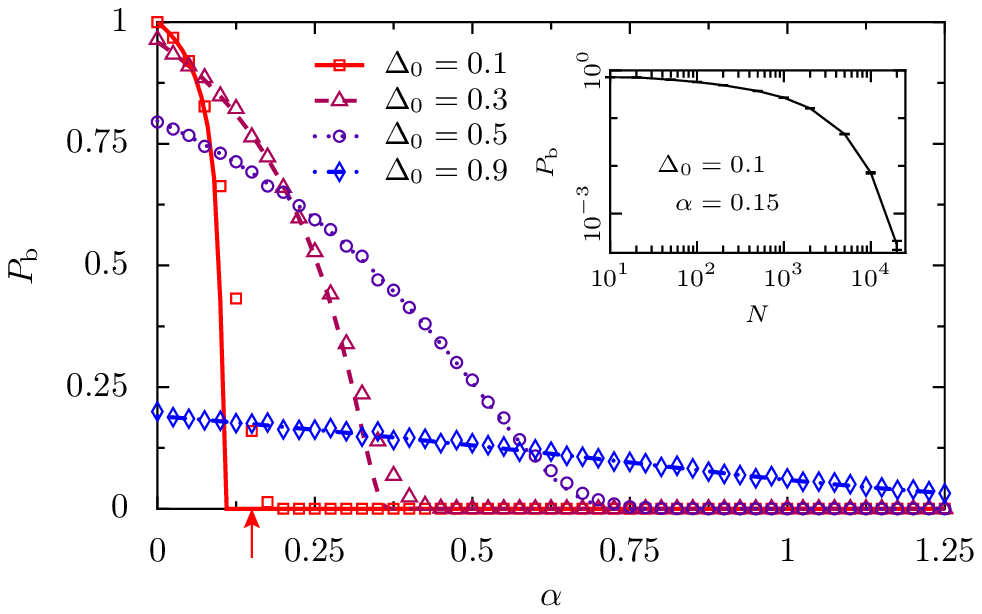}
  \caption{No-cascade (upper panel) and breakdown (lower panel)
    probabilities, $\Pnc$ and $\Pb$, respectively, as a function of
    the  tolerance~$\alpha$ for different values of the
    load-redistribution parameter~$\Delta_0$. Lines: Eq.~\eqref{eq:Pnc} (upper panel) and 
    results from an iterative solution of
    Eq.~\eqref{eq:pb_ex} (lower panel). Symbols: Monte-Carlo results from a
    simulation of the stochastic dynamics~\eqref{eq:redist_1} for
    $N=2000$ elements and $10^4$ realizations. The statistical error
    is below the size of the symbols. Inset of lower panel:
    Monte-Carlo results (from $10^5$ realizations) as a function of
    the system size~$N$ for $\Delta_0=0.1$ and $\alpha=0.15$, as
    indicated by the arrow in the lower panel. The line serves as a
    guide to the eye.}
  \label{fig:pb_alpha}
\end{figure}

Figure~\ref{fig:pb_alpha} shows the  probabilities $\Pnc$ and $\Pb$ as a
function of the tolerance~$\alpha$ for various values of the
redistribution factor~$\Delta_0$.
We observe (see upper panel) that the no-cascade probability~$\Pnc$
gradually increases from its minimal value $\exp(-1/\Delta_0)$ for
$\alpha=0$ but remains considerably below one over the considered
$\alpha$-range. This is in stark contrast to the behavior of the
breakdown probability (see lower panel), which decreases with
increasing tolerance~$\alpha$ to vanish completely above a certain
critical $\alpha$-value. In this latter regime, the system becomes
stable in the sense that cascading failures affecting it as a whole do
not occur with finite probability. With increasing load-redistribution
factor~$\Delta_0$, the transition to this regime happens at higher
$\alpha$-values and also becomes less sharp. Comparing with the
no-cascade probabilities~$\Pnc$, we find that those cannot serve as a
reliable indicator for the system stability: Consider, for example,
the case $\alpha=0.5$, where the breakdown probability varies strongly
with $\Delta_0$, as opposed to the no-cascade probability, which is
even independent of $\Delta_0$ for $\Delta_0\le\alpha$.

In Fig.~\ref{fig:pb_alpha}, we also compare the results from the
generalized-branching-process approximation with those obtained from
a Monte-Carlo simulation of the full stochastic
dynamics~\eqref{eq:redist_1} for a system consisting of $N=2000$
elements (see symbols in Fig.~\ref{fig:pb_alpha}). Within the
statistical error, we find a very good agreement, except near the
transition to a stable system in the case of small
load-redistribution factors $\Delta_0$. In this regime, the failing
load is distributed to a large number of elements, but not all of them
fail immediately. Their increased load, however, will eventually lead
to a higher breakdown probability than predicted by the
branching-process approximation, where this effect is neglected.  As
the number of such elements is independent of the system size, this
finite-size effect will vanish in the limit of very large systems. As
shown exemplarily for the case~$\Delta_0=0.1$ and $\alpha=0.15$ in the
inset of Fig.~\ref{fig:pb_alpha}, the breakdown probability obtained
from Monte-Carlo simulations indeed approaches zero with increasing
system size~$N$, in agreement with the solution obtained from Eq.~\eqref{eq:pb_ex}.

\begin{figure}[t]
  \centering
  \includegraphics[width=0.96\linewidth]{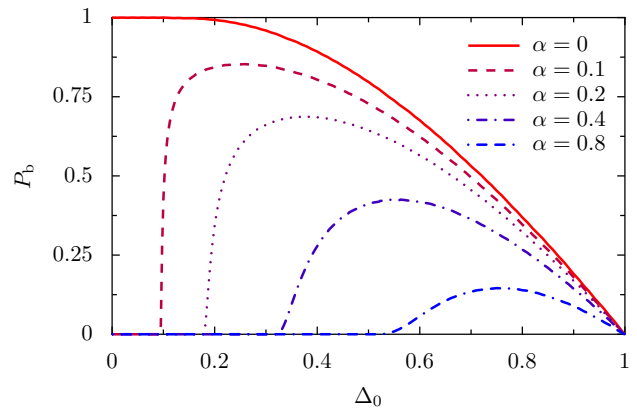}
  \caption{Breakdown robability~$\Pb$ obtained from an
    iterative solution of Eq.~\eqref{eq:pb_ex} as a function of the
    load-redistribution parameter~$\Delta_0$ for different values of
    the element tolerance~$\alpha$. }
  \label{fig:pb_delta0}
\end{figure}
It is also interesting to look at the behavior of the breakdown
probability as a function of the load-redistribution factor~$\Delta_0$
(see Fig.~\ref{fig:pb_delta0}). For a fixed tolerance~$\alpha$, we
find a vanishing breakdown probability~$\Pb$ for small $\Delta_0$,
which corresponds to ``well-connected'' systems where the failing load
is redistributed to a large number of other elements. Above a critical
$\Delta_0$-value, the breakdown probability increases abruptly, in
particular for small tolerances $\alpha$. It reaches a maximum and
then gradually decreases again towards zero in the limit of $\Delta_0$
going to unity, where the failing load is transferred to a single
other element. It follows that the network is robust against cascading
breakdown if $\Delta_0$ is smaller than its $\alpha$-dependent
critical value.

Finally, we compare our results with those of a simple branching
process model, e.g., Refs.~\cite{Dobson2007,Dobson2005}, where the
induced load increments are independent of the load of the failing
element.  In these models, the no-cascade probability $\Pnc$ as well
as the breakdown probability $\Pb$ are completely determined by a
single quantity, the mean number $\mu\f$ of failures that are induced
by a failing element.  In particular, $\Pb$ is zero if $\mu\f<1$ and
finite if $\mu\f>1$.  When such a model is applied to our
  prototype example, the load increments after a failure are equal to
  a constant $Q_0$ with probability $p_0$ and zero otherwise. It
  follows that $\mu\f = \min(1/2\alpha, 1/2Q_0)$, and we note that for
  a consistent comparison with our model, $Q_0$ has to be identified
  with $\Delta_0/2$. As a function of $\alpha$, the breakdown
  probability $\Pb$ then becomes zero at the critical
  value~$\alpha_\mathrm{c} = 1/2$, which is independent of the value
  of $\Delta_0$. In contrast to our results of
  Fig.~\ref{fig:pb_delta0}, we thus find that such a model does not
  exhibit a critical behavior with respect to the
  parameter~$\Delta_0$, i.e., the breakdown probability stays finite
  for arbitrarily small values of $\Delta_0$ if $\alpha <1/2$. For
  $\alpha \ge 1/2$, $\Pb$ is zero for all values of $\Delta_0$ ($0 \le
  \Delta_0 \le 1$).

\section{Conclusions}

We have introduced and analyzed a class of stochastic
failure-propagation models which, compared to previous approaches,
enable a more realistic description of real systems, while still being
amenable to an analytical treatment. The approach is applied to a
prototype example that is motivated by the propagation of line
failures in power transmission networks. The initial loads (power
flows) have random values and the maximum load an element can carry is
assumed to be equal to $(1 + \alpha)$ times its initial load,
Eq.~\eqref{eq:Lcrit}. With this example, we have demonstrated that our
model not only exhibits a critical behavior as a function of the
failure tolerance~$\alpha$, but also with respect to a
parameter~$\Delta_0$ that characterizes the variance of the
load-redistribution factors and thus depends on physical as well as on
topological properties of the load or flow dynamics.

While our assumption of stochastic load redistribution neglects any
spatial correlations, we are still able to gain new 
insights into the vulnerability of complex networks. If we use a more
realistic distribution of redistribution factors~$\Delta$, our results
on the critical behavior of the breakdown probability with respect to
failure tolerance and “connectivity”, e.g., may give valuable
information for the design of more robust infrastructure systems.

Finally, we note that our models can not only be applied to critical
infrastructure networks, but also to other breakdown phenomena, e.g.,
to failure propagation in elastic fiber
bundles~\cite{Alava2006,Pradhan2008}. Within our approach, a
corresponding model (with stochastic load redistribution) is obtained
if we assume that the initial loads are all identical and that the
critical loads of the individual elements are randomly distributed.  A
detailed study of such models will be presented in a separate
publication~\cite{Lehmann2010}.

\section*{Acknowledgments}

We thank X.\ Feng and J.D.\ Finney for fruitful discussions.

\end{document}